\documentclass[review,12pt]{elsarticle}

\biboptions{sort&compress}
\usepackage{amssymb}
\usepackage{amsmath}
\usepackage{hyperref}
\usepackage{float}
\usepackage{comment}
\usepackage{makecell}
\usepackage{xcolor}
\begin{document}
\sloppy
\begin{frontmatter}



\title{Selective recovery of critical materials in supercritical water desalination}


\author{Tae Jun Yoon\corref{cor1}}
%\ead{tyoon@lanl.gov}
\author{Erica P. Craddock}
\author{Jeremy C. Lewis}
\author{John A. Matteson}
\author{Jong Geun Seong}
\author{Rajinder P. Singh}
\author{Katie A. Maerzke}
\author{Robert P. Currier}
\author{Alp T. Findikoglu}
\address{Los Alamos National Laboratory, Los Alamos, NM 87545, United States}
\cortext[cor1]{tyoon@lanl.gov}

\begin{abstract}
	Supercritical water desalination (SCWD) is an alternative zero-liquid discharge desalination technique that can overcome many technical and environmental challenges in common desalination processes. Our recent techno-economic analyses on the process integration and intensification of an SCWD process suggest that SCWD can be an economically competitive zero-liquid discharge desalination technique for highly concentrated brines. In addition to these attractive features, this work explores the possibility of utilizing the SCWD process for the selective recovery of industrially essential materials as co-products. Model brines containing sodium chloride, neodymium chloride, and other sodium-containing salts, are examined as model feeds from 25 to 450 $^\circ$C at 25 MPa. The sodium contents in the effluent were not sensitive to the presence other salts, but that of neodymium was. When sodium acetate was added, water-insoluble precipitates were obtained. The characterization of these deposits using gas pycnometer, FT-IR, SEM/EDS, and XRD indicates that the precipitates mainly contain neodymium hydroxide and some low-concentration impurities. These results suggest that an SCWD process has the potential to recover critical materials while producing freshwater.
\end{abstract}

\begin{comment}
\begin{graphicalabstract}
	\includegraphics{GraphicalAbstract.eps}
\end{graphicalabstract}

\onecolumn
\begin{highlights}
	\item Model brines containing neodymium and sodium salts are desalinated in the SCWD process.
	\item A theoretical framework based on DEW and MSA models is suggested for ion speciation.
	\item Highly concentrated NaCl in water facilitates NdCl\textsubscript{3} precipitation.
	\item Ion association at high temperatures is strongly affected by the types of anions.
	\item Nd(OH)\textsubscript{3} was obtained as a main insoluble precipitate in acetate-containing systems.
\end{highlights}
\end{comment}
%\twocolumn

\end{frontmatter}


\section{Introduction}
Desalination, a physical/chemical process for removing salts from brine, is a feasible route for addressing water scarcity. The ever-increasing need for drinkable water has led to the advancement of many desalination technologies \cite{curto2021review,zheng2014seawater} such as reverse osmosis \cite{qasim2019reverse}, multi-stage flash \cite{el1999multi}, multiple-effect distillation \cite{zheng2014seawater}, mechanical vapor compression \cite{zejli2011optimization}, and electrodialysis \cite{al2020electrodialysis}. These techniques are relatively mature and show good performance for potable water production, but one of their major drawbacks is that a highly concentrated brine is discharged from the plant. Since the concentrated brine can harm the environment \cite{ruso2007spatial}, an additional discharge step (e.g., surface discharge, deep well injection, and evaporation pond) or process (e.g., brine concentrator and crystallizer) is required \cite{giwa2017brine,zhang2020reverse,ariono2016brine}. To avoid this expense, zero-liquid discharge techniques that do not produce any liquid wastes are being explored \cite{nakoa2016sustainable,yaqub2019zero}.

Supercritical water desalination (SCWD) is one option for realizing zero-liquid discharge desalination \cite{odu2015design,van2018design,van2020analysis,van2020potential,yaqub2019zero,sharan2021energy}. In this process, a salt-containing aqueous mixture is pressurized and heated above the critical point of pure water ($T_\mathrm{c}=374\ ^\circ$C and $p_\mathrm{c}=22.1$ MPa). The typical operating condition in the SCWD process is, in fact, not \textit{beyond the critical point of the brine}; the critical temperature and pressure in concentrated salt solutions are much higher than that of pure water. Above a threshold temperature, the compressed brine is split into a distillate and bottoms fraction. The salt concentration in the distillate is usually less than the potable water limit ($<1,200$ ppm), and the remaining salts are concentrated in the bottoms fraction. The bottom stream is usually rapidly depressurized to obtain additional clean water as a superheated vapor and precipitate the salts.

The SCWD process offers other advantages compared to commercial desalination processes. Above all, it accommodates a variety of feeds consisting of highly concentrated brines (well above the seawater levels) and wastewater containing organic pollutants. Common organic substances can be quickly decomposed into CO\textsubscript{2} and H\textsubscript{2}O at high temperatures via exothermic oxidation reactions \cite{bermejo2006supercritical,marrone2013supercritical,veriansyah2005supercritical}.

On the other hand, the high thermal energy input required to reach operating conditions is a significant obstacle impeding the widespread commercialization of SCWD processes. This disadvantage can be partially avoided by using waste heat from power plants or by process integration. Our recent techno-economic analyses demonstrated that the energy use and the water cost could be significantly reduced by integrating SCWD with other processes \cite{sharan2021energy}. For example, the heat generated by the oxidation of organic pollutants or hydrocarbons can even generate energy beyond that necessary to operate the process \cite{van2020potential,cocero2002supercritical,loppinet2008current,queiroz2015supercritical}.

This work explores another potential advantage of SCWD. Specifically, the possibility of selective recovery of critical materials that are commercially or strategically important is explored. Industrial wastewater \cite{kawasaki1998rare}, acid mine drainage \cite{zhao2007geochemistry,ayora2016recovery,vital2018treatment}, groundwater in the vicinity of mine sites \cite{yan2013geochemistry}, and produced water \cite{tian2020rare} are known to contain critical materials that are indispensable in high-tech industries, such as lithium, nickel, copper, scandium, yttrium, and lanthanides (rare earth elements, or REEs). Since both the feed composition and operating conditions control the solubility of salts in near-critical and supercritical water \cite{voisin2017solubility}, an SCWD process might work synergetically for recovering such precious elements while producing drinkable water.

We designed, built, and executed a series of experiments involving supercritical desalination of model solutions in a continuous flow unit. The operating temperature varied from 20 to 450 $^\circ$ C along an isobar (25 MPa). The model feed solutions consisted of neodymium chloride (NdCl\textsubscript{3}), sodium chloride (NaCl), different sodium salts, and distilled water. By analyzing the distillate compositions and precipitates and combining them with theoretical calculations, we demonstrate that SCWD shows considerable promise for the selective recovery of critical materials while producing potable water.

\section{Methods}
\subsection{Materials}
Distilled water (H\textsubscript{2}O, Kroger\textsuperscript{\tiny\textregistered}) was locally purchased. The specific conductance of the distilled water was determined as $<1\mathrm{\mu S/cm}$ using a handheld conductometer (EW-19601-03, Cole-Parmer\textsuperscript{\tiny\textregistered}). All salts, including sodium chloride (Sigma Aldrich, NaCl, $\geq99.0\ \%$), sodium sulfate (Sigma Aldrich, Na\textsubscript{2}SO\textsubscript{4}, $\geq99.0\ \%$), sodium acetate (Sigma Aldrich, CH\textsubscript{3}COONa or Na(Ac), $\geq99.0\ \%$), and neodymium chloride hexahydrate (Sigma Aldrich, $\mathrm{NdCl\textsubscript{3}\cdot(H_2O)_6}$, $\geq99.0\ \%$), were purchased from Sigma Aldrich. No additional purification was performed on the substances.
\subsection{SCWD apparatus}
Before each experiment, feed solutions were prepared by weighing the salts (Model MS 204S/03, Mettler Toledo\textsuperscript{\tiny\textregistered}) and adding them to 2 L of distilled water (see Table \ref{tab1:feed-composition} for the feed compositions). The concentration of neodymium in solutions was selected to be approximately 200 ppm (based on Nd\textsuperscript{3+}), considering the lowest detection limit of the ultraviolet-visible spectrophotometer.
\begin{table}
	\caption{Model feed compositions studied in this work. Each run was named as $b$ (binary), $c$ (chloride), $s$ (sulfate), and $a$ (acetate).}
	\begin{center}
	\begin{tabular}{ccccc}
		\hline
		Run & Anion (A) & \makecell{$c_\mathrm{NaCl}^0$\\(ppm)} & \makecell{$c_\mathrm{NdCl_3}^0$\\(ppm)} & \makecell{$c_\mathrm{Na_xA_y}^0$\\ (ppm)}\\
		\hline
		$\mathrm{b1}$ & None & 10,168 & 0 & 0 \\
		$\mathrm{b2}$ & None & 0 & 344 & 0 \\
		$\mathrm{c1}$ & None & 1,067 & 347 & 0 \\
		$\mathrm{c2}$ & None & 10,168 & 332 & 0 \\
		$\mathrm{c3}$ & None & 104,547 & 334 & 0 \\
		$\mathrm{s1}$ & $\mathrm{SO_4^{2-}}$ & 9,924 & 334 & 590 \\
		$\mathrm{s2}$ & $\mathrm{SO_4^{2-}}$ & 9,681 & 349 & 1,181 \\
		$\mathrm{s3}$ & $\mathrm{SO_4^{2-}}$ & 8,952 & 347 & 2,954 \\
		$\mathrm{a1}$ & $\mathrm{CH_3COO^-}$ & 9,679 & 347 & 685 \\
		$\mathrm{a2}$ & $\mathrm{CH_3COO^-}$ & 9,190 & 343 & 1,371 \\
		$\mathrm{a3}$ & $\mathrm{CH_3COO^-}$ & 7,737 & 400 & 3,412 \\
		\hline
	\end{tabular}
	\end{center}
	\label{tab1:feed-composition}
\end{table}

\begin{figure*}
	\includegraphics[width=\textwidth]{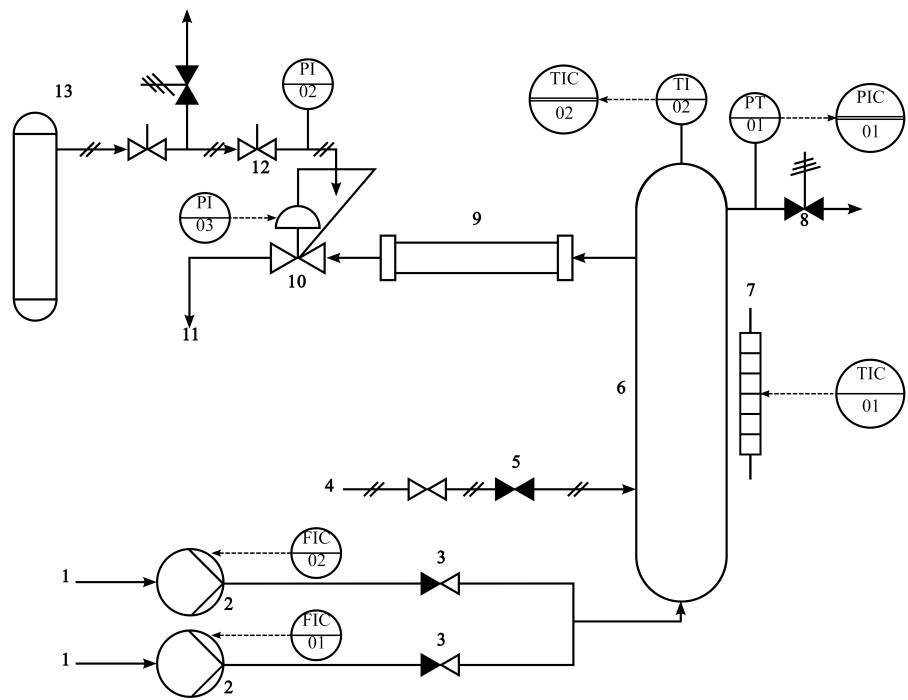}
	\caption{Schematic of the supercritical water desalination (SCWD) unit designed and operated in this work: 1, model brine (feed); 2, high-pressure liquid chromatography (HPLC) pump; 3, in-line check valve; 4, purge gas; 5, ball valve; 6, vessel; 7, electrical heater; 8, pressure relief device; 9, single-pass heat exchanger; 10, dome-type back pressure regulator (BPR) connected to a forward pressure regulator; 11, distillate (effluent); 12, needle valve; 13, 6000 psi argon cylinder.}
	\label{fig1: process-diagram}
\end{figure*}
	Figure \ref{fig1: process-diagram} shows the process flow diagram of the continuous desalination unit. The unit is identical to the \textit{in-situ} conductometric device described in our earlier work \cite{yoon2021insitu}. In this work, the \textit{in-situ} electrochemical sensor was removed since the precipitates generated in concentrated brine feeds could contaminate the electrodes. 
	
	In this unit, two high-pressure LS Class pumps (LS-
	Class Pump 903034 REV L, Teledyne SSI) supply brine solution(s) from the bottom into a tubular vessel. A split-type heater (A3653HC20EE, ThermCraft\textsuperscript{\tiny\textregistered}) can heat the vessel to 500 $^\circ$C. Two thermocouples are used to measure and control the system temperature; one of the thermocouples measures the solution temperature, and the other is located outside the vessel and measures the vessel surface temperature. After passing through the vessel, the solution is cooled by a single-pass heat exchanger that uses water as a coolant (RTE-111, Neslab\textsuperscript{\tiny\textregistered}). The operating pressure is regulated by a back pressure regulator (BPR, A93VB, Parr Instrument Company\textsuperscript{\tiny\textregistered}) connected to a 40 MPa (6,000 psi) Argon cylinder ($\geq$ 99.999 \%, Matheson
	Tri-Gas\textsuperscript{\tiny\textregistered}). The pressure of the BPR dome was regulated by a spring-loaded forward pressure regulator (FPR, 44-1166-24,
	Tescom\textsuperscript{\tiny\textregistered}). Additional details are described elsewhere \cite{yoon2021insitu}.
	
	The basic operating procedure was as follows. Before every experiment, the vessel was flushed with distilled water at 300 $^\circ$C and 10 MPa for 3 hours. Then, distilled water was continuously flowed for 3 hours to remove any impurities or corrosion products. The vessel was dried at 120 $^\circ$C overnight. Then, the desalination vessel (c.a. 420 mL) was filled with the sample by supplying feed solution at 20 mL/min. One liter of the feed solution was supplied to the vessel to remove any remaining water or impurities within the unit. Then, the feed flow rate was decreased to 3 mL/min, and the system pressure was increased to 25 MPa. When the system reached the target pressure, the temperature increased from 25 $^\circ$C to 450 $^\circ$C. Approximately 10 mL of the effluent was collected when the system reached the target temperature for compositional analysis.
	\begin{figure}
		\begin{center}
		\includegraphics[width=0.5\textwidth]{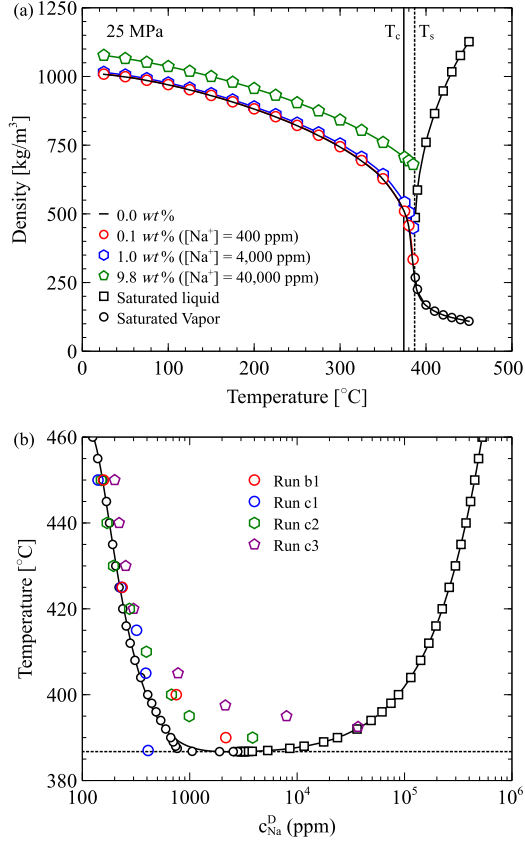}
		\caption{(a) Bulk solution densities as a function of temperature and feed concentration calculated from the Driesner correlation. (b) NaCl (\textit{aq}) concentrations in vapor and liquid phases above the phase separation temperature ($T_\mathrm{s}$). Open circles and squares denote the vapor and liquid branches, respectively. The colored open symbols are the distillate concentrations obtained in this work (see Table and main text for the detailed description of the feed composition in each run). The total concentration of sodium in the distillates ($c_\mathrm{Na}^\mathrm{D}$) was slightly higher than those from the Driesner model below 420 $^\circ$C. This discrepancy mainly occurs through either carryovers or some remaining salt solutions in the cooling section.}
		\label{fig2:phase-diagrams}
		\end{center}
	\end{figure}

	Figure \ref{fig2:phase-diagrams} (a) shows the bulk densities of NaCl (\textit{aq}) at 25 MPa as a function of temperature and NaCl concentrations calculated based on the Driesner correlation \cite{driesner2007system1,driesner2007system2}. The densities of pure water were calculated from the Wagner-Pru\ss~Equation of State (EoS) \cite{wagner2002iapws}. All NaCl solutions studied in this work ($w_\mathrm{NaCl}<10$ \textit{wt}\% where $w$ is the weight fraction) exist as homogeneous compressed liquids below the vapor-liquid separation temperature ($T_\mathrm{s}=386.74\ ^\circ$C). Figure \ref{fig2:phase-diagrams} (b) shows the NaCl partition between vapor and liquid above $T_\mathrm{s}$. The phase coexistence lines were also calculated from the Driesner correlation.
	
	After the experiment was completed, the vessel was cooled and depressurized. The cooled vessel was then disassembled and examined. When insoluble precipitates were observed inside the vessel, they were collected by washing the inner parts with distilled water and scraping the vessel's bottom head and wall. The collected precipitates were filtered, washed and sonicated in distilled water, and dried overnight in an oven at 125 $^\circ$C.	
	
\subsection{Analytical procedures}
	A sodium-selective electrode (LAQUAtwin Na-11 ion meter, Horiba\textsuperscript{\tiny\textregistered}) was used to measure the total concentration of sodium in the distillate (effluent). The sodium-selective electrode was calibrated with standard solutions (150 ppm and 2,000 ppm) every day. The electrode resolution was 1 ppm ($0-99$ ppm), 10 ppm ($100-990$ ppm), and 100 ppm ($1,000-9,900$ ppm), and its accuracy level was $\pm10$ \%, according to the manufacturer. For each sample, the measurements were performed in triplicate.
	
	An ultraviolet-visible spectrophotometer (Evolution 201, Thermo Scientific, Inc.) was used to determine the total neodymium concentration in the effluent ($c_\mathrm{Nd}^\mathrm{D}$). When neodymium was added to the sample, a strong absorption was observed at wavelengths around 574 nm. Thus, the calibration curve was constructed by subtracting the minimum absorbance strength (baseline) from the maximum absorbance around 574 nm (see the ultraviolet spectra and calibration curve in the Supplementary Material). The resultant calibration curve showed good linearity in the concentration range from 20 to 4,000 ppm (based on the Nd\textsuperscript{3+} concentration). Below 20 ppm, the height of the absorption peak became indistinguishable from the background noise. The calibration curve was also tested against various samples that contain other organic/inorganic species to check the interference effect. No significant change was observed in the prominent peak of the ultraviolet spectra and the readings from the sodium ion meter.
	
	In order to characterize the precipitates obtained in the SCWD process, skeletal density measurements, optical and scanning electron microscopy (SEM), energy-dispersive X-ray spectroscopy (EDS), Fourier transform infrared spectroscopy (FT-IR), and X-ray diffraction analysis (XRD) were conducted. Skeletal density measurements were conducted on precipitates with a gas pycnometer (AccuPyc II 1345, Micromeritics\textsuperscript{\tiny\textregistered}) using helium as the analysis gas. The densities were measured repeatedly, since some impurities (e.g., grahite gasket) from the reactors were mixed with the samples. All samples were held in a $0.1\ \mathrm{cm^3}$ sample cell.
	
	Field-emission scanning electron microscopy (FE-SEM, FEI Inspect\textsuperscript{TM}) and energy-dispersive X-ray spectroscopy (EDS) techniques were employed to observe the morphology of the precipitates and confirm the presence of neodymium. The samples were prepared by dispersing the precipitates on a carbon tape adhered to an aluminum stub. Precipitates were coated with gold (c.a. 5 nm) using a sputter coater (Q150R, Quorum Technologies, Ltd.) to maximize X-ray counts during EDS and minimize charging effects. A roughly 5 min. EDS scan at low resolution was adequate to show the presence of neodymium in the precipitates with a high degree of certainty. The precipitate color was demonstrated by imaging the crystals with a digital optical microscope (VHX-7000, Keyence). The precipitate was imaged on a reflective stage using white, combined ring and confocal lighting at $300\times$ zoom. A digital image was taken through the microscope lens with the fully integrated head (VHX-7100, Keyence).
	
	ATR-FTIR analysis was performed with a PIKE Technologies VeeMAX III ATR accessory on a Thermo Fisher Nicolet 6700 FTIR spectrometer (resolution: $2\ \mathrm{cm}^{-1}$, 64 scans). The single bounce Zn/Se crystal temperature was maintained at 35 $^\circ$C with a temperature controller (GladiATR\textsuperscript{TM}, PIKE Technologies). The samples were clamped with $0.2\ \mathrm{N{\cdot}m}$ of torque to ensure contact with the crystal. Since some parts of the graphite gaskets detached from the desalination unit were mixed with the precipitates, the baseline slope had to be corrected. The baseline correction and the signal-to-noise ratio enhancement were performed based on the correction algorithm by Zhang et al. \cite{zhang2010baseline} and the Savitzky-Golay filter \cite{savitzky1964smoothing}.

	Wide angle X-ray diffraction (XRD) analysis was performed to investigate crystal structure of the precipitates. A diffractometer with a graphite diffracted beam monochromator (Siemens D5000) was used for the investigation and Cu K$\alpha$ (1.54 \AA) was a source for the radiation. The XRD spectra were obtained at 2$\theta$ angle range from 5$^\circ$ to 60$^\circ$ with a scanning speed of 0.4$^\circ$/ min. The detailed experimental procedure was also presented in the previous report by Zhao et al. \cite{zhao2020graphite}. The measured spectra were analyzed by comparing them with theoretical spectra calculated from the unit cell structures. The unit cell structures were obtained from Crystallography Open Database (COD) \cite{gravzulis2012crystallography} and Inorganic Crystal Structure Database (ICSD) \cite{allmann2007introduction}.
	
	\subsection{Theoretical analysis}
	To elucidate the behavior of electrolyte solutions at elevated temperatures and pressures, we performed ion speciation calculations at different temperatures along the 25 MPa isobar. When cation(s) $\mathrm{C}$ and anion(s) $\mathrm{A}$ form an ion pair $\mathrm{P}$, the chemical equilibrium between these species is represented as:
	\begin{equation}
		\nu_\mathrm{C}\mathrm{C}+\nu_\mathrm{A}\mathrm{A}\rightleftharpoons\mathrm{P}
		\label{eq:speciation-simple}
	\end{equation}
	where $\nu_\mathrm{i}$ denotes the stoichiometric coefficients of the species $i$ forming the ion pair $\mathrm{P}$. The ion association constant of this \textit{reaction} $K_\mathrm{a,c}$ in molarity scale is given as:
	\begin{equation}
		K_\mathrm{a,c}=\frac{a_\mathrm{P}}{a_\mathrm{A}^\mathrm{\nu_A}a_\mathrm{B}^\mathrm{\nu_B}}=\frac{c_\mathrm{P}}{c_\mathrm{A}^\mathrm{\nu_A}c_\mathrm{B}^\mathrm{\nu_B}}f_\mathrm{P}^{-1}
	\end{equation}
	where $a_\mathrm{i}$ is the activity of the species $i$, a product of the concentration $c_\mathrm{i}$ and the activity coefficient $y_\mathrm{i}$. The activity coefficient ratio is denoted as $f_\mathrm{P}$. The molarity-based association constant was obtained by converting the molality-based constant ($K_\mathrm{a,m}$) following the formula given by Robinson and Stokes \cite{robinson2002electrolyte}. In the conversion from molality to molarity, solution density is required. Since the feed concentration of sodium chloride was much higher than the other salts, we assumed that the density of the mixed electrolyte solution was essentially equal to the density of NaCl (\textit{aq}) calculated from the Driesner model.
	\begin{figure}
		\begin{center}
		\includegraphics[width=0.7\textwidth]{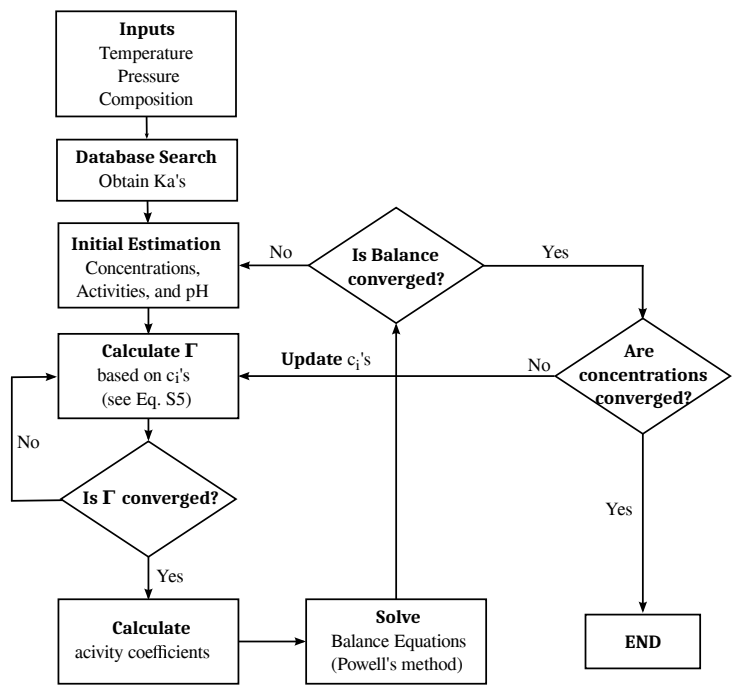}
		\caption{Chemical speciation algorithm designed in this work. The DEWPython library was connected to the mean spherical approximation (MSA) algorithm to obtain the association constants and activity coefficients. Chemical equilibria together with the mass and charge balance equations were solved by using the Newton-Krylov method. For the detailed description, see the Supplementary Material. $\Gamma$ denotes the screening parameter, a key parameter in the MSA calculation. }
		\label{fig3: speciation-algorithm}
		\end{center}
	\end{figure}

	Figure \ref{fig3: speciation-algorithm} shows the ion speciation calculation procedure used in this work. In this procedure, ion association constants and activity coefficients are first calculated. Most ion association constant data were obtained from the modified Helgeson-Kirkham-Flowers (HKF) model, also known as the Deep Earth Water (DEW) model \cite{zimmer2016supcrtbl,sverjensky2014water,shock1992calculation,shock1993metal,zhang2005prediction,fernandez1997formulation,delany1978calculation}. The DEWPython module \cite{chan2021dewpython}, a python-based software to calculate the thermodynamic properties in high-temperature brines, was connected to the activity coefficient calculation algorithm. DEW model provides a general thermodynamic framework to predict a wide range of geophysical properties. However, the predicted properties disagree with experimental measurements in some substances. For instance, the first and the second ion association constants between neodymium and chloride ions from experiments and MD simulations \cite{finney2019ion,stepanchikova1999spectrophotometric,gammons1996aqueous,migdisov2002spectrophotometric} are considerably different from those predicted by the DEW parameters proposed by Haas et al. \cite{haas1995rare}. Thus, when the discrepancy is considerable, or no HKF parameter is available (e.g., $\mathrm{Nd(SO_4)_2}^{-1}$ \cite{wood1990aqueous,migdisov2006spectrophotometric} and $\mathrm{(Na_2SO_4)}^0$ \cite{hnedkovsky2005electrical}), empirical results were chosen by comparing literature (See the Supplementary Material for the detailed description).
	
	A variety of activity coefficient models have been proposed, including Debye-H\"uckel equation, Helgeson equation \cite{helgeson1974theoretical}, Bromley equation \cite{bromley1972approximate}, Pitzer equation \cite{pitzer1974thermodynamics}, and mean spherical approximation (MSA) \cite{blum1975mean}. We used the MSA model to calculate the activity coefficients since it will also be used to estimate thermophysical properties from conductance measurements \cite{bernard1992conductance} in our future work. The MSA model dissects the activity coefficient into hard-sphere ($y_\mathrm{i}^\mathrm{hs}$) and electrostatic contributions ($y_\mathrm{i}^\mathrm{el}$) as $\ln{y_\mathrm{i}}=\ln{y_\mathrm{i}^\mathrm{hs}}+\ln{y_\mathrm{i}^\mathrm{el}}$. Each contribution was calculated based on hard-sphere radii, charge, and number density of ionic species (see the Supplementary Material for the detailed calculation procedure for the activity coefficients). Crystallographic radii taken from Marcus \cite{marcus1988ionic} were used as the hard-sphere radii. In calculating the electrostatic part, the Bjerrum radius (instead of the hard-sphere radius) was used, following the suggestion by Sharygin et al. \cite{sharygin2001tests}. Since the activity coefficients in the MSA framework depend on the concentration of charged species in a system, an iterative procedure was required.
	
	After determining the activity coefficients and the association constant, the ion speciation calculation was conducted by solving the mass balance and the electroneutrality condition (charge balance). For instance, the concentration of each species in Eq. (2) can be obtained by solving the following set of equations.
	\begin{subequations}
		\begin{equation}
			c_\mathrm{C}^0=c_\mathrm{C}+\nu_\mathrm{C}c_\mathrm{P}
		\end{equation}
		\begin{equation}
			c_\mathrm{A}^0=c_\mathrm{A}+\nu_\mathrm{C}c_\mathrm{P}
		\end{equation}
		\begin{equation}
			c_\mathrm{P}=K_\mathrm{a}f_\mathrm{P}c_\mathrm{C}^\mathrm{\nu_C}c_\mathrm{A}^\mathrm{\nu_A}
		\end{equation}
	\end{subequations}
	Although this simple example is analytically solvable, the ion speciation procedure in our system is much more complex since (1) neodymium ion shifts the pH by making hydroxide complexes, and (2) it makes multiple ion pairs. For instance, in ``simple'' NdCl\textsubscript{3} (\textit{aq}), at least nine chemical equilibria should be considered.
	\begin{subequations}
		\begin{equation}
			\mathrm{Nd^{3+}}+n\mathrm{Cl^{-}}\rightleftharpoons{(\mathrm{NdCl}_n)^{3-n}}
		\end{equation}
		\begin{equation}
			\mathrm{Nd^{3+}}+n\mathrm{(OH)^{-}}\rightleftharpoons{[\mathrm{Nd(OH)}_n]^{3-n}}
		\end{equation}
		\begin{equation}
			\mathrm{H^{+}}+\mathrm{(OH)^{-}}\rightleftharpoons\mathrm{H_2O}
		\end{equation}
	\end{subequations}
	where the stoichiometric coefficient $n$ varies from one to four. Note that we neglected the formation of mixed ion pairs (e.g., ${\mathrm{Nd(OH)}_m\mathrm{(Cl)}_n}$) due to the absence of ion association constants. Nevertheless, the concentration of twelve ionic species should be solved simultaneously. The conjugated direction method proposed by Powell \cite{powell1964efficient} was adopted to solve this set of equations. Since the solution does not converge easily \cite{brassard2000feasible}, different initial estimates were attempted for each system. When a stable solution was obtained, the calculation was repeated until the concentrations of all species converged. Considering that a description of the calculation details and all relevant assumptions is quite lengthy, all procedures are described in detail in the Supplementary Material.
	
	\section{Results and Discussion}
	\begin{figure*}
		\begin{center}
		\includegraphics[width=\textwidth]{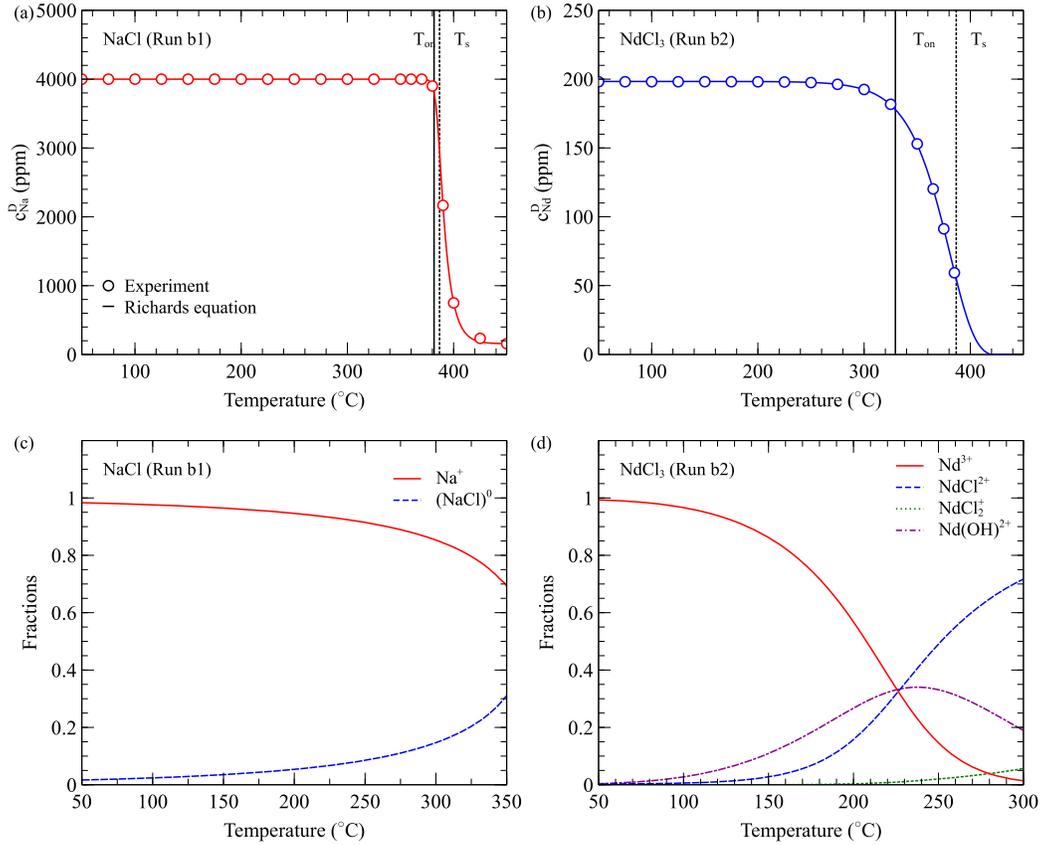}
		\caption{(a) Sodium ($c_\mathrm{Na}^\mathrm{D}$) and (b) neodymium ($c_\mathrm{Nd}^\mathrm{D}$) concentrations in the effluents (distillates) as a function of the operating temperature in binary systems. The onset temperatures ($T_\mathrm{on}$) were obtained as 381.6 (Run $\mathrm{b1}$) and 329.4 $^\circ$C (Run $\mathrm{b2}$), respectively. The onset temperature in Run $\mathrm{b1}$ is close to the phase separation temperature $T_\mathrm{s}$, suggesting that the decrease in $c_\mathrm{Na}^\mathrm{D}$ mainly originates from the phase separation. Ion speciation in these systems [(c) and (d)] demonstrates that the temperature dependence of the ion association behaviors in these systems is also completely different.}
		\label{fig4: binary}
		\end{center}
	\end{figure*}
	
	Figure \ref{fig4: binary} shows the sodium and neodymium concentrations in effluents ($c_\mathrm{Na}^\mathrm{D}$ and $c_\mathrm{Nd}^\mathrm{D}$) as a function of temperature in the Runs $\mathrm{b1}$ and $\mathrm{b2}$. In both runs, only a single salt was dissolved in the distilled water. No appreciable decrease in $c_\mathrm{Na}^\mathrm{D}$ was observed in the sodium chloride solution ($c_\mathrm{Na}^0=4,000$ ppm) until the system temperature reached 385 $^\circ$C. When the operating temperature was beyond that threshold temperature, the volumetric flow rate of the effluent increased significantly, and the $c_\mathrm{Na}^\mathrm{D}$ profile showed a sharp decrease. This behavior suggests that the $c_\mathrm{Na}^\mathrm{D}$ profile change is almost purely attributed to vapor/liquid separation in the solution. According to the Driesner correlation, the phase separation in NaCl (\textit{aq}) systems at 25 MPa occurs at 386.74 °C (Figure \ref{fig2:phase-diagrams}). The effluent concentrations obtained from different experiments (Runs $\mathrm{b1}$, $\mathrm{c1}$, $\mathrm{c2}$, and $\mathrm{c3}$) agree well with the vapor composition line calculated from the Driesner correlation, except for data very near the threshold temperature. These results suggest that the operating conditions with the desalination unit are close to equilibrium. The discrepancies observed near the phase separation temperature likely arise from liquid remaining in the single-pass heat exchanger and/or back-pressure regulator or possibly carry-over of the concentrated brine from the vessel.
	
	The temperature dependence of the concentration profiles along the isobar is well described by a generalized logistic equation (Richards equation \cite{richards1959flexible}), which is expressed as:
	\begin{equation}
		\frac{c_\mathrm{i}-c_\mathrm{i}^\mathrm{min}}{c_\mathrm{i}^0-c_\mathrm{i}^\mathrm{min}}=\frac{1}{\left[1+\exp\left[a(T-T_\mathrm{thr})\right]\right]^{b}}
		\label{eq:richards}
	\end{equation}
	where $c_\mathrm{i}$ is the concentration of the species $i$ in ppm, and $c_\mathrm{i}^0$ is the feed concentration. Other quantities $c_\mathrm{i}^\mathrm{min}$, $a$, $b$, and $T_\mathrm{thr}$ are adjustable parameters. The onset temperature $T_\mathrm{on}$ where the concentration profile starts to decrease steeply was calculated from Eq. \ref{eq:richards}. From tedious but straightforward algebra, it is obtained as:
	\begin{equation}
		T_\mathrm{on}=T_\mathrm{thr}+\frac{2}{ab}(1-2^{b})
	\end{equation}		
	where $a$ and $b$ are the adjustable parameters in Eq. \ref{eq:richards}. In Run $\mathrm{b1}$ (NaCl), $T_\mathrm{on}$ was obtained as 381.6 $^\circ$C, close to the phase separation temperature. Thus, NaCl separation should be driven only by phase separation. On the other hand, the onset temperature in Run $\mathrm{b2}$ (NdCl\textsubscript{3}) was calculated as 329.4 $^\circ$C. Although the onset temperature for neodymium chloride is significantly lower than that of sodium chloride, it should be noted that a considerable amount of neodymium ($c_\mathrm{Nd}^\mathrm{D}=59.3$ ppm) still exists in the effluent at 385 $^\circ$C.
	
	Ion speciation analysis on these binary systems suggests that their ion association behaviors are entirely different. In Run $\mathrm{b1}$, no appreciable ion pair formation occurs up to 243 $^\circ$C; thermal/kinetic energy can easily overcome the ion aggregation induced by the decrement in the dielectric constant. Above 243 $^\circ$C, a decrease in the dielectric constant of water stabilizes ion pairs. The temperature where the free ion concentration starts to decrease is close to the threshold temperature found using molecular dynamics (MD) simulations and separate experimental measurements \cite{yoon2019electrical,yoon2021insitu}. On the other hand, the free ion concentration in Run $\mathrm{b2}$ showed an almost sigmoidal dependence on temperature. It starts to decrease near 150 $^\circ$C. Interestingly, neodymium hydroxide [$\mathrm{Nd(OH)}^{2+}$] and neodymium monochloride [$\mathrm{NdCl}^{2+}$] are formed as dominating species below 300 $^\circ$C. The formation of neodymium hydroxide complex suggests that multiple ion aggregation (e.g., ${\mathrm{Nd}^{3+}+x\mathrm{Cl}^-+y\mathrm{(OH)}^-}$), which could not be considered in the theoretical calculation due to the lack of the association constants, would play an important role in decreasing $c_\mathrm{Nd}^\mathrm{D}$; the decrease in $c_\mathrm{Nd}^\mathrm{D}$ observed in the experiment is significantly affected by the pH level \cite{han2019effect}.	
	\begin{figure*}
		\begin{center}
		\includegraphics[width=\textwidth]{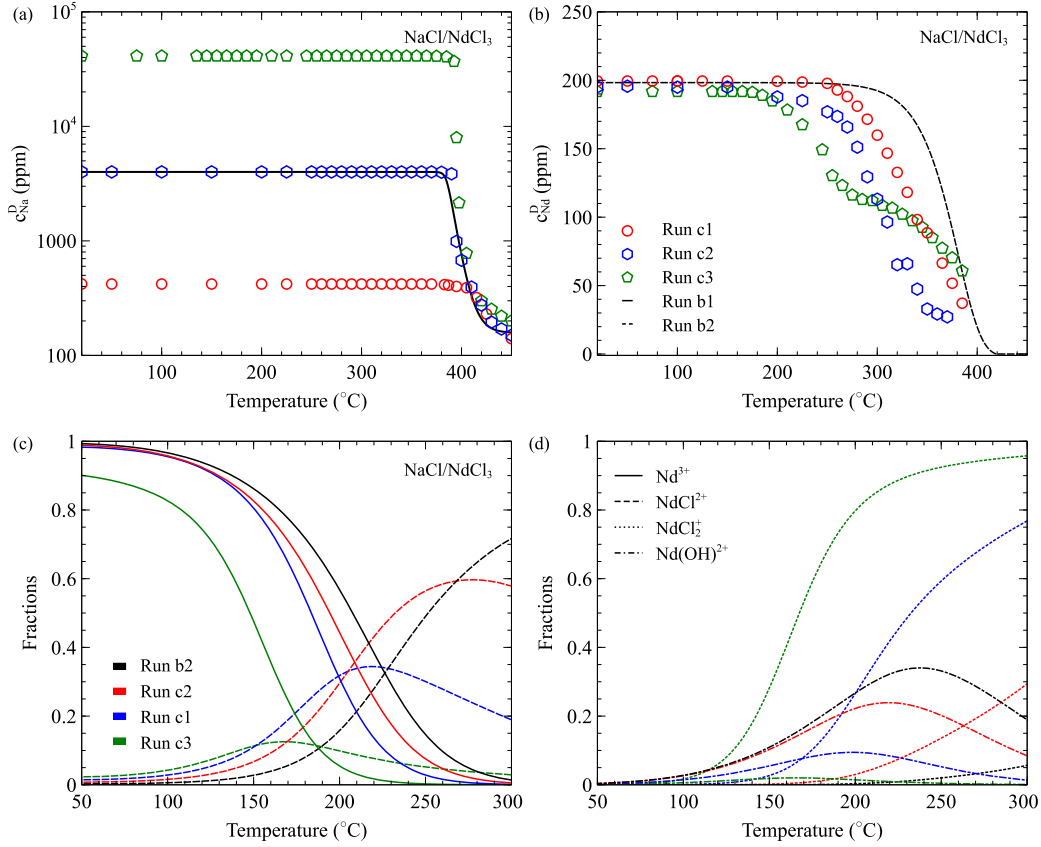}
		\caption{The concentration of (a) sodium ($c_\mathrm{Na}^\mathrm{D}$) and (b) neodymium ($c_\mathrm{Nd}^\mathrm{D}$) in the distillate as a function of the NaCl feed concentration and temperature in the NaCl/NdCl\textsubscript{3} mixtures (25 MPa). Despite the significant change in the NaCl feed concentration, the normalized sodium concentration in the effluent ($c_\mathrm{Na}^\mathrm{D}/c_\mathrm{Na}^0$) shows an almost identical dependence on temperature. This result suggests that the separation of sodium ions would be driven by the phase change. On the contrary, $c_\mathrm{Nd}^\mathrm{D}$ shows a complex temperature/concentration dependence. Ion speciation analyses in (c) and (d) demonstrate that ion pair formation between $\mathrm{Nd}^{3+}$ and $\mathrm{Cl^-}$ is facilitated as $c_\mathrm{Na}^0$ increases. The presence of $\mathrm{Nd^{3+}}$ does not have a significant influence on the formation of $\mathrm{(NaCl)^0}$.}
		\label{fig5: NaCl-NdCl3}
		\end{center}
	\end{figure*}

	Next, we examine the influence of NaCl concentration in the feed solution (Figure \ref{fig5: NaCl-NdCl3} (a) and (b), Runs $\mathrm{c1-c3}$). Although $c_\mathrm{Na}^\mathrm{0}$ was changed one hundredfold (from 420 to 41,128 ppm) and neodymium chloride was added in the feed, $c_\mathrm{Na}^\mathrm{D}$ did not show a significant change. This result again confirms that sodium chloride precipitation is mainly driven by the vapor-liquid separation at 386.74 $^\circ$C. On the contrary, $c_\mathrm{Nd}^\mathrm{D}$ profile was notably affected by $c_\mathrm{Na}^0$. As $c_\mathrm{Na}^0$ increased, the onset temperature of $\mathrm{Nd}^{3+}$ decreased from 277.5 (Run $\mathrm{c1}$) to 255.7 (Run $\mathrm{c2}$) to 191.8 $^\circ$C (Run $\mathrm{c3}$). Moreover, $c_\mathrm{Nd}^\mathrm{D}$ showed a double-sigmoidal behavior in Run $\mathrm{c3}$. It decreased sharply at the first onset temperature, decreased slowly between 191.8 to 310 $^\circ$C, and then decreased sharply again. 
	Ion speciation analysis results [Figure \ref{fig5: NaCl-NdCl3} (c) and (d)] explain the decrease in $T_\mathrm{on}$. As the chloride ion concentration increases, the chemical equilibrium between $\mathrm{Nd^{3+}}$ and $\mathrm{Cl^-}$ is shifted, facilitating the formation of ion pairs. As a result, $\mathrm{NdCl_2^+}$ becomes a dominating species among Nd-containing ion aggregates. 
	
	There are two possible contributions to the double-sigmoidal behavior in $c_\mathrm{Nd}^\mathrm{D}$ profile (Run $\mathrm{c3}$) at high temperatures. First, as given in Figure \ref{fig2:phase-diagrams} (a), the solution density in Run $\mathrm{c3}$ is much higher that in other Runs. High solution density results in the solubility increase of neutral species not considered in the numerical calculation (e.g., ${\mathrm{Nd(OH)}_n\mathrm{Cl}_{(3-n)}}$). Second, it should be noted that the dielectric constant of pure water was only used in this work. Indeed, the presence of NaCl in aqueous media can \textit{increase} the dielectric constant in high-temperature brines. In a highly polar medium like ambient water, the addition of salts typically decreases the dielectric constant, which arises from kinetic depolarization \cite{maribo2013modeling}. In a low-dielectric medium, the formation of multiple ion aggregates enhances the mean dipole moment in a system \cite{wang2001computation}. Our recent MD simulations on high-temperature NaCl solutions modeled with several different molecular models also support the idea that the dielectric constant of high-temperature brine is increased as the brine concentration increases at simulated pressures of 20 MPa (approximately 30 MPa in reduced units) \cite{patel2021nacl}. According to the MD simulations, a dielectric constant increment is observed when the temperature is above the critical temperature of pure water. This dielectric enhancement will also increase the solubility of free neodymium ions and charged ion pairs. The overall results suggest that the presence of NaCl can alter the precipitation behavior of NdCl\textsubscript{3}. This phenomenon has great potential to be utilized in controlling the selective recovery of neodymium from highly concentrated brines.	
	\begin{figure*}
		\begin{center}
			\includegraphics[width=\textwidth]{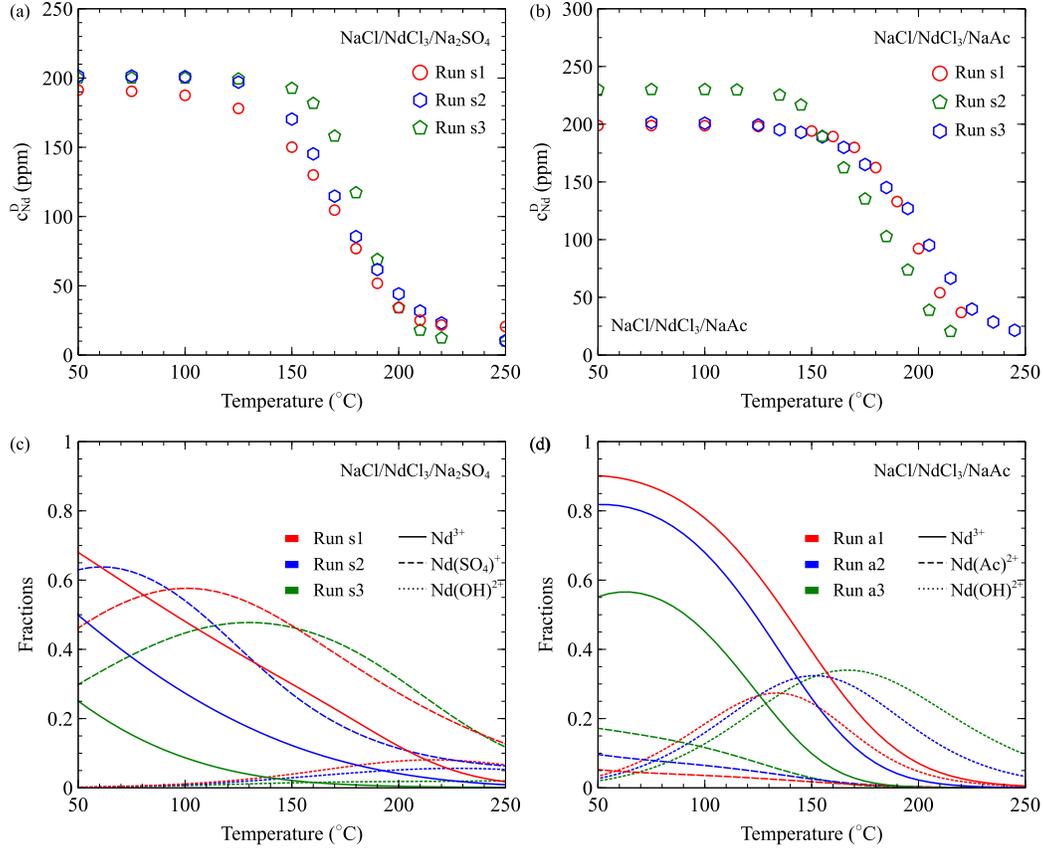}
			\caption{Concentration of neodymium ions in (a) NaCl/NdCl\textsubscript{3}/Na\textsubscript{2}SO\textsubscript{4} systems (Runs $\mathrm{s1-s3}$) and (b) NaCl/NdCl\textsubscript{3}/NaAc systems (Runs $\mathrm{a1-a3}$). In the sulfate systems, the onset temperature of neodymium slightly increased as the sodium sulfate concentration in the feed solution increased. This behavior reflects the retrograde solubility of sulfate salts in the system. On the contrary, the onset temperature in the acetate systems decreased as the sodium acetate concentration increased. Ion speciation analyses given in (c) and (d) show that the ion aggregation tendency in water is $\mathrm{Cl^-<Ac^-\ll{SO_4^{2-}}}$ below the onset temperatures, as demonstrated in room temperature experiments \cite{friesen2018hydration,han2019effect}. Note that the fraction of hydroxide complex $\mathrm{Nd(OH)^{2+}}$ is very high in acetate systems, which explains the formation of water-insoluble precipitates in these Runs. For other species, see the Supplementary Material.}
			\label{fig6: effect-of-anion}
		\end{center}
	\end{figure*}

	Next, we examined the influence of different anions on neodymium ion concentration in the effluent. Figure \ref{fig6: effect-of-anion} (a) and (b) shows the concentration of neodymium ions in NaCl/NdCl\textsubscript{3}/Na\textsubscript{2}SO\textsubscript{4} systems (Runs $\mathrm{s1-s3}$) and in NaCl/NdCl\textsubscript{3}/NaAc systems (Runs $\mathrm{a1-a3}$). Consistent with the previous operations, sodium concentrations in the effluent were not affected by the presence of neodymium or other anions (see the numerical data in the Supplementary Material). On the other hand, the onset temperature of neodymium in sulfate-containing systems was altered significantly. $T_\mathrm{on}^\mathrm{sulfate}$ was obtained as 131.0, 141.2, and 162.6, $^\circ$C as the sodium sulfate concentration in the feed increased. That is, the addition of sulfate increased the onset temperature slightly. This behavior could come from the retrograde solubility of neodymium sulfate, which is commonly observed in sulfate systems \cite{migdisov2006spectrophotometric,van2019calcium}. Even though the sulfate salts show retrograde behavior, the average onset temperature in NaCl/NdCl\textsubscript{3}/Na\textsubscript{2}SO\textsubscript{4} systems was quite a bit lower (145 $^\circ$C) compared to non-sulfate systems. It is even lower than that obtained from Run c3 ($T_\mathrm{on}^\mathrm{c3}=191.8\ ^\circ$C). 
	
	These results can also be understood from the ion speciation analyses [Figure \ref{fig6: effect-of-anion} (c)]. When sulfate exists, $c_\mathrm{Nd}$ shows a steep decrease with increasing temperature, and $\mathrm{(NdSO_4)^+}$ becomes a dominating species below the onset temperatures. The concentrations of chloride and hydroxide complexes were much lower than those in NaCl/NdCl\textsubscript{3} systems.
	
	When sodium acetate (NaAc) was added, the onset temperature changed from 169.1 (Run $\mathrm{a1}$) to 161.0 (Run $\mathrm{a2}$) to 143.2 $^\circ$C (Run $\mathrm{a3}$). The onset temperature was decreased as the sodium acetate concentration increased, and no retrograde behavior was observed. The average onset temperature was higher than that of the sulfate systems. In the ion speciation analysis, the fractions of free neodymium ions in the acetate systems were higher than those in the sulfate systems but lower than those in the chloride systems below the onset temperatures. Similar to the sulfate systems, only a small amount of neodymium chloride complexes (see Figure S4) were observed. However, the dominating species near the onset temperatures were \textit{not neodymium acetate but neodymium hydroxide complexes}. The fraction of hydroxide complexes is significantly increased due to the pH shift induced by sodium acetate. Since acetic acid is an associating acid, hydrogen ions in water make $\mathrm{(HAc)^0}$ complexes while increasing the pH level. The overall results suggest that the ion aggregation tendency is $\mathrm{Cl^-<Ac^-\ll{SO_4^{2-}}}$, which agrees with earlier studies at ambient conditions \cite{friesen2018hydration,han2019effect}.
	
	Although the neodymium-acetate ion pair formation is not as prevalent as neodymium-sulfate ion association, we did observe that insoluble precipitates were obtained after all acetate runs (Runs $\mathrm{a1-a3}$), whereas no insoluble precipitates were obtained in the other experiments. Three types of reactions are likely responsible for the generation of water-insoluble precipitates. First, thermal decarboxylation of sodium acetate yields methane and carbon dioxide in hydrothermal environments according to:
	\begin{equation}
		\mathrm{2CH\textsubscript{3}COONa+H\textsubscript{2}O\rightarrow2CH\textsubscript{4}+Na\textsubscript{2}CO\textsubscript{3}+CO\textsubscript{2}}
	\end{equation}
	Carbon dioxide generated in this reaction can make ion pairs with the dissolved neodymium ions. The decarboxylation/oxidation mechanism and role of experimental factors, including reactor surface, catalyst, temperature, pressure, pH, and feed concentration, have been studied extensively \cite{palmer1986thermal,bell1994thermal,mccollom2003experimental,onwudili2010hydrothermal,ong2013situ,li2017experimental}. There are some variations in the threshold temperature for this reaction, but it is generally accepted that the decomposition readily occurs above 400 $^\circ$C. In this route, double carbonate salt ($\mathrm{NaNd(CO_3)_2}$) (instead of neodymium carbonate $\mathrm{Nd_2(CO_3)_3}$) is known to be preferred \cite{mochizuki1974synthesis,rao1999solubility,yang2019metastable}. Considering that $c_\mathrm{Nd}^\mathrm{D}$ decreased significantly below 250 $^\circ$C, this reaction mechanism would not significantly contribute to the formation of insoluble precipitates. Another possible mechanism is the thermal degradation of neodymium acetate. When exposed to high temperatures, neodymium acetate [Nd(Ac)\textsubscript{3}] is degraded to neodymium monohydroxide acetate (Nd(OH)(Ac\textsubscript{3})\textsubscript{2}), $334-345\ ^\circ$C), neodymium monoxide carbonate (Nd\textsubscript{2}O(CO\textsubscript{3})\textsubscript{2}), $359-370\ ^\circ$C), neodymium dioxide carbonate (Nd\textsubscript{2}O\textsubscript{2}CO\textsubscript{3}), $371-410\ ^\circ$C) and neodymium oxide (Nd\textsubscript{2}O\textsubscript{3}, $684-700\ ^\circ$C) \cite{shaplygin1979thermogravimetric,kkepinski2004hydrothermal}. Lastly, hydrothermal synthesis of neodymium hydroxide [Nd(OH)\textsubscript{3}] or hydroxide-containing crystals (e.g., Nd(OH)\textsubscript{2.45}(Ac)\textsubscript{0.55} \cite{kkepinski2004hydrothermal} or Nd(OH)\textsubscript{2}Cl \cite{bukin1972neutron,zehnder2010investigation}) would be feasible, depending on the reaction environment.
	
	Together with the ion speciation analyses above [Figure \ref{fig6: effect-of-anion} (d)], we characterized the precipitate to elucidate what types of water-insoluble precipitates were obtained in Runs $\mathrm{a1-a3}$. The skeletal densities of our samples were obtained as $\rho_\mathrm{sk}=4.674\pm0.084\ \mathrm{g/cm^3}$ (Run $\mathrm{a1}$), $\rho_\mathrm{sk}=4.398\pm0.364$ $\mathrm{g/cm^3}$ (Run $\mathrm{a2}$), and $\rho_\mathrm{sk}=4.892\pm0.344$ $\mathrm{g/cm^3}$ (Run $\mathrm{a3}$). Theoretical skeletal densities of some neodymium crystals were obtained either directly from literature or by calculating the X-ray densities from the unit cell structures. They are given as $\rho_\mathrm{sk}^\mathrm{Nd(Ac)_3}$=2.179 $\mathrm{g/cm^3}$ \cite{gomez2008anhydrous}, $\rho_\mathrm{sk}^\mathrm{Nd(OH)_2(Cl)}$=4.730 $\mathrm{g/cm^3}$ \cite{bukin1972neutron}, $\rho_\mathrm{sk}^\mathrm{Nd(OH)_3}$=4.777 $\mathrm{g/cm^3}$ \cite{beall1976refinement},
	$\rho_\mathrm{sk}^\mathrm{Nd(OH)CO_3}$=4.510 $\mathrm{g/cm^3}$ \cite{dexpert1974determination}, $\rho_\mathrm{sk}^\mathrm{Nd_2O_2(CO)_3}$=6.331 $\mathrm{g/cm^3}$ \cite{christensen1970hydrothermal}, and $\rho_\mathrm{sk}^\mathrm{Nd_2O_3}$=7.24 $\mathrm{g/cm^3}$. Assuming that adsorbed water molecules do not significantly influence the measurement results, the density of the obtained precipitates is close to hydroxide-containing crystals.
	
	\begin{figure*}
		\begin{center}
			\includegraphics[width=\textwidth]{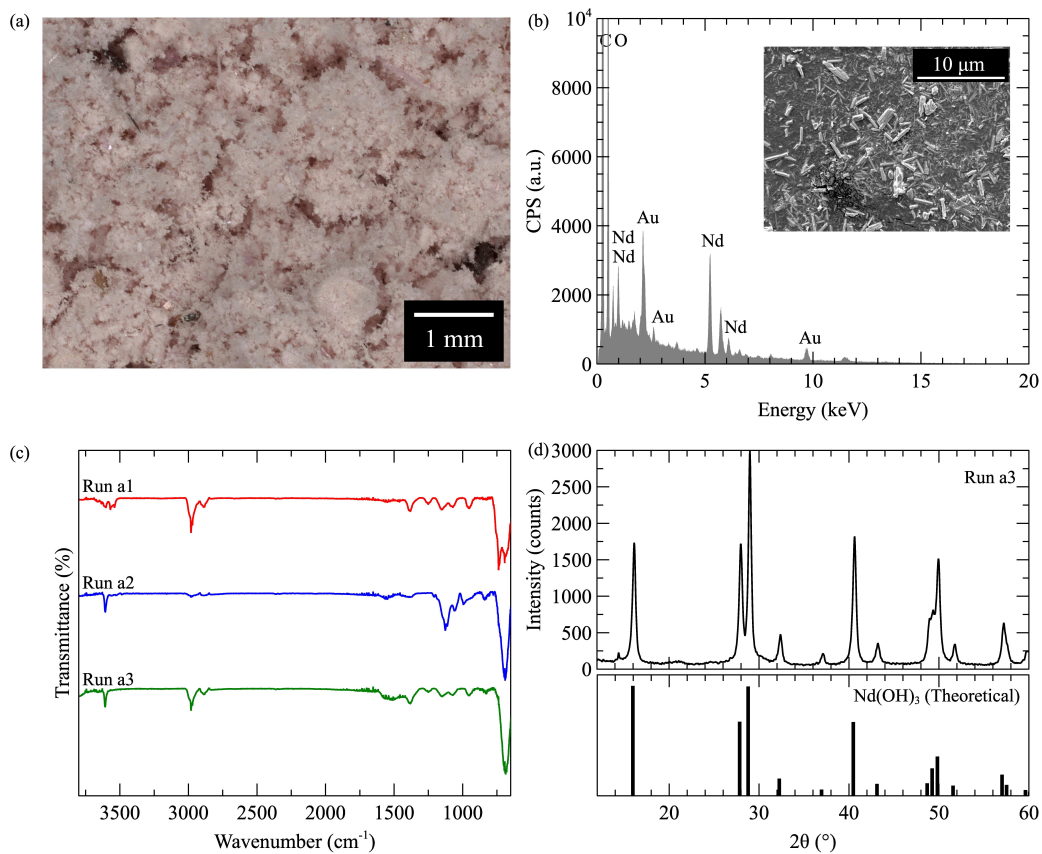}
			\caption{Characterization of the precipitates obtained in Run $\mathrm{a1-a3}$. (a) Optical spectroscopy image of the precipitate from Run $\mathrm{a1}$. The precipitate color is pinkish. (b) SEM-EDS analyses on the samples obtained in Run $\mathrm{a2}$. The energy-dispersive X-ray spectroscopy reveals that the water-insoluble precipitates contain neodymium. (c) Infrared (IR) spectra of the neodymium-bearing precipiates from Run $\mathrm{a3}$. The IR spectrum suggests that the precipitate contains hydroxide group and neodymium. (d) X-ray diffraction spectrum of the precipitate obtained in Run $\mathrm{a3}$ (top) and the theoretical XRD spectrum calculated from the unit cell structure of Nd(OH)\textsubscript{3} (bottom). The consistency between the spectra suggests that the precipitate mainly consists of neodymium and hydroxide (with some impurities).}
			\label{fig7:characterization-precipitates}
		\end{center}
	\end{figure*}
	Figure \ref{fig7:characterization-precipitates} (a) and (b) shows the appearance of the obtained precipitates. The precipitates were pinkish to the naked eye and under white light, which is a general color observed in neodymium salts. SEM/EDS analysis revealed that the obtained crystals definitely do contain neodymium and have a rod-like shape. Figure \ref{fig7:characterization-precipitates} (c) shows the FT-IR spectra of the precipitates. They suggest that the precipitate contains the hydroxide group ($\mathrm{OH}$), which shows a sharp absorption peak at 3610 $\mathrm{cm}^{-1}$. A strong peak around 680 $\mathrm{cm}^{-1}$ corresponds to the Nd-OH interaction, which is observed in oxygen-containing neodymium salts such as neodymium acetate \cite{kkepinski2004hydrothermal}. Many weak signals were observed within 950-1600 $\mathrm{cm^{-1}}$, which might be relevant to the carboxyl group (1560 $\mathrm{cm^{-1}}$) and the methyl group (1153 $\mathrm{cm^{-1}}$) observed in $\mathrm{Nd(Ac)_3}$ crystals \cite{shaplygin1979thermogravimetric}. However, they could not be clearly distinguished since the signal strength was relatively weak. Figure \ref{fig7:characterization-precipitates} (d) shows the X-ray diffraction spectrum of the precipitates from Run $\mathrm{a3}$. In all samples, the prominent peaks in the spectrum match with those observed in the neodymium hydroxide (see the Supplementary Material). The X-ray diffraction spectra of other potential substances (e.g., Nd\textsubscript{2}CO\textsubscript{3} and Nd\textsubscript{2}O\textsubscript{3}) did not show a complete match with that of the precipitate. Similar to the FT-IR spectra, the diffraction spectra in other samples were similar except that there were some shifts in peak positions and non-identifiable peaks. Considering all characterization results, the obtained precipitates would mainly consist of neodymium hydroxide with some low-concentration anions such as acetate, carbonate, and chloride. K{\c e}pi{\'nski}, Zawadzki and Mi{\`s}ta, for instance, obtained Nd(OH)\textsubscript{2.45}(Ac)\textsubscript{0.55}$\cdot$0.45H\textsubscript{2}O crystals by a hydrothermal reaction of neodymium acetate \cite{kkepinski2004hydrothermal}. Overall, this result suggests that the selective recovery of neodymium from concentrated brine solutions in a water-insoluble precipitate form would be feasible by adding acetate.
	
	\section{Conclusions}
	In this work, a series of supercritical water desalination experiments were conducted on model brines that consist of water, sodium chloride, neodymium chloride, sodium acetate, and sodium sulfate from 20 to 450 $^\circ$C along an isobar of 25 MPa. 
	
	The distillate concentration of sodium chloride, one of the most common salts in nature, shows a drastic decrease when vapor-liquid separation occurs at 387 $^\circ$C. NaCl distribution between the distillate and bottoms was not affected significantly by other salts or their feed concentration. This result suggests that vapor-liquid separation is primarily responsible for the sodium chloride separation in the SCWD process.
	
	In contrast, neodymium (one of the most sought-after rare earth elements) shows a noticeable dependence on the brine composition. The onset temperature where the neodymium concentration in the effluent decreases is reduced dramatically when the sodium chloride concentration is high or small amounts of sulfate or acetate salts exist in the feed. Ion speciation analysis suggests that ion pair formation of the neodymium with anions is responsible for this change.
	
	When sodium acetate is added to the brine, a water-insoluble precipitate is obtained. Characterization of the precipitates suggest that this insoluble precipitate is predominantly neodymium hydroxide with some impurities such as acetate, chloride, and carbonate.
	
	This work examined whether supercritical water desalination can be utilized for selective recovery of critical materials while simultaneously producing potable water. The experimental and theoretical calculations suggest that it has a considerable potential to retrieve critical materials by way of either liquid adducts (concentrate) or water-insoluble salts. Taking into account that (1) supercritical water desalination is insensitive to organic impurities in the feed and (2) the concentration of critical materials in wastewater or some natural water sources can be relatively high (ppm levels), it is expected that the supercritical desalination with co-product production may prove to be an attractive option for economically viable and environmentally benign zero-liquid discharge desalination. It will be interesting to study feed solutions containing other naturally common salts or industrially essential elements to further test the applicability of the supercritical water desalination process with selective recovery of valuable components. 
	
	\section*{Acknowledgement}
	This work was supported by the Director’s Postdoctoral Fellow Program (20190653PRD4) and the Laboratory Directed Research and Development Program (20190057DR) at Los Alamos National Laboratory.
	
	\section*{Declaration of Competing Interest}
	The authors declare that they have no known competing financial interests or personal relationships that could have influenced the work reported in this paper.
	
	\section*{CRediT authorship contribution statement}
	\textbf{Tae Jun Yoon:} Conceptualization; Methodology; Software; Validation; Formal analysis; Investigation; Data Curation; Writing - Original Draft; Writing - Review \& Editing; Visualization; Funding acquisition
	\textbf{Erica P. Craddock:} Formal analysis; Writing - Review \& Editing
	\textbf{Jeremy C. Lewis:} Formal analysis; Writing - Review \& Editing
	\textbf{John A. Matteson:} Formal analysis; Writing - Review \& Editing
	\textbf{Jong Geun Seong:} Formal analysis; Writing - Review \& Editing
	\textbf{Rajinder P. Singh:} Resources; Writing - Review \& Editing
	\textbf{Katie A. Maerzke:} Writing - Review \& Editing; Funding acquisition
	\textbf{Robert P. Currier:} Writing - Review \& Editing; Supervision; Project administration; Funding acquisition
	\textbf{Alp T. Findikoglu:} Writing - Review \& Editing; Supervision; Funding acquisition
\bibliographystyle{elsarticle-num}
\bibliography{bibliography}

\begin{thebibliography}{10}
\expandafter\ifx\csname url\endcsname\relax
  \def\url#1{\texttt{#1}}\fi
\expandafter\ifx\csname urlprefix\endcsname\relax\def\urlprefix{URL }\fi
\expandafter\ifx\csname href\endcsname\relax
  \def\href#1#2{#2} \def\path#1{#1}\fi

\bibitem{curto2021review}
D.~Curto, V.~Franzitta, A.~Guercio, A review of the water desalination
  technologies, Applied Sciences 11~(2) (2021) 670.

\bibitem{zheng2014seawater}
X.~Zheng, D.~Chen, Q.~Wang, Z.~Zhang, Seawater desalination in china:
  Retrospect and prospect, Chemical engineering journal 242 (2014) 404--413.

\bibitem{qasim2019reverse}
M.~Qasim, M.~Badrelzaman, N.~N. Darwish, N.~A. Darwish, N.~Hilal, Reverse
  osmosis desalination: A state-of-the-art review, Desalination 459 (2019)
  59--104.

\bibitem{el1999multi}
H.~T. El-Dessouky, H.~M. Ettouney, Y.~Al-Roumi, Multi-stage flash desalination:
  present and future outlook, Chemical Engineering Journal 73~(2) (1999)
  173--190.

\bibitem{zejli2011optimization}
D.~Zejli, A.~Ouammi, R.~Sacile, H.~Dagdougui, A.~Elmidaoui, An optimization
  model for a mechanical vapor compression desalination plant driven by a
  wind/pv hybrid system, Applied energy 88~(11) (2011) 4042--4054.

\bibitem{al2020electrodialysis}
S.~Al-Amshawee, M.~Y. B.~M. Yunus, A.~A.~M. Azoddein, D.~G. Hassell, I.~H.
  Dakhil, H.~A. Hasan, Electrodialysis desalination for water and wastewater: A
  review, Chemical Engineering Journal 380 (2020) 122231.

\bibitem{ruso2007spatial}
Y.~D.~P. Ruso, J.~A. De~la Ossa~Carretero, F.~G. Casalduero, J.~S. Lizaso,
  Spatial and temporal changes in infaunal communities inhabiting soft-bottoms
  affected by brine discharge, Marine environmental research 64~(4) (2007)
  492--503.

\bibitem{giwa2017brine}
A.~Giwa, V.~Dufour, F.~Al~Marzooqi, M.~Al~Kaabi, S.~Hasan, Brine management
  methods: Recent innovations and current status, Desalination 407 (2017)
  1--23.

\bibitem{zhang2020reverse}
X.~Zhang, Y.~Liu, Reverse osmosis concentrate: an essential link for closing
  loop of municipal wastewater reclamation towards urban sustainability,
  Chemical Engineering Journal (2020) 127773.

\bibitem{ariono2016brine}
D.~Ariono, M.~Purwasasmita, I.~G. Wenten, Brine effluents: Characteristics,
  environmental impacts, and their handling., Journal of Engineering \&
  Technological Sciences 48~(4) (2016).

\bibitem{nakoa2016sustainable}
K.~Nakoa, K.~Rahaoui, A.~Date, A.~Akbarzadeh, Sustainable zero liquid discharge
  desalination (szldd), solar Energy 135 (2016) 337--347.

\bibitem{yaqub2019zero}
M.~Yaqub, W.~Lee, Zero-liquid discharge (zld) technology for resource recovery
  from wastewater: A review, Science of the total environment 681 (2019)
  551--563.

\bibitem{odu2015design}
S.~O. Odu, A.~G. van~der Ham, S.~Metz, S.~R. Kersten, Design of a process for
  supercritical water desalination with zero liquid discharge, Industrial \&
  Engineering Chemistry Research 54~(20) (2015) 5527--5535.

\bibitem{van2018design}
S.~van Wyk, S.~O. Odu, A.~G. van~der Ham, S.~R. Kersten, Design and results of
  a first generation pilot plant for supercritical water desalination (scwd),
  Desalination 439 (2018) 80--92.

\bibitem{van2020analysis}
S.~van Wyk, A.~G. van~der Ham, S.~R. Kersten, Analysis of the energy
  consumption of supercritical water desalination (scwd), Desalination 474
  (2020) 114189.

\bibitem{van2020potential}
S.~van Wyk, A.~G. van~der Ham, S.~R. Kersten, Potential of supercritical water
  desalination (scwd) as zero liquid discharge (zld) technology, Desalination
  495 (2020) 114593.

\bibitem{sharan2021energy}
P.~Sharan, J.~D. McTigue, T.~J. Yoon, R.~Currier, A.~T. Findikoglu, Energy
  efficient supercritical water desalination using a high-temperature heat
  pump: A zero liquid discharge desalination, Desalination 506 (2021) 115020.

\bibitem{bermejo2006supercritical}
M.~Bermejo, M.~Cocero, Supercritical water oxidation: a technical review, AIChE
  journal 52~(11) (2006) 3933--3951.

\bibitem{marrone2013supercritical}
P.~A. Marrone, Supercritical water oxidation—current status of full-scale
  commercial activity for waste destruction, The Journal of Supercritical
  Fluids 79 (2013) 283--288.

\bibitem{veriansyah2005supercritical}
B.~Veriansyah, T.-J. Park, J.-S. Lim, Y.-W. Lee, Supercritical water oxidation
  of wastewater from lcd manufacturing process: kinetic and formation of
  chromium oxide nanoparticles, The Journal of supercritical fluids 34~(1)
  (2005) 51--61.

\bibitem{cocero2002supercritical}
M.~Cocero, E.~Alonso, M.~Sanz, F.~Fdz-Polanco, Supercritical water oxidation
  process under energetically self-sufficient operation, The Journal of
  Supercritical Fluids 24~(1) (2002) 37--46.

\bibitem{loppinet2008current}
A.~Loppinet-Serani, C.~Aymonier, F.~Cansell, Current and foreseeable
  applications of supercritical water for energy and the environment,
  ChemSusChem: Chemistry \& Sustainability Energy \& Materials 1~(6) (2008)
  486--503.

\bibitem{queiroz2015supercritical}
J.~Queiroz, M.~Bermejo, F.~Mato, M.~Cocero, Supercritical water oxidation with
  hydrothermal flame as internal heat source: Efficient and clean energy
  production from waste, The Journal of Supercritical Fluids 96 (2015)
  103--113.

\bibitem{kawasaki1998rare}
A.~Kawasaki, R.~Kimura, S.~Arai, Rare earth elements and other trace elements
  in wastewater treatment sludges, Soil Science and Plant Nutrition 44~(3)
  (1998) 433--441.

\bibitem{zhao2007geochemistry}
F.~Zhao, Z.~Cong, H.~Sun, D.~Ren, The geochemistry of rare earth elements (ree)
  in acid mine drainage from the sitai coal mine, shanxi province, north china,
  International Journal of Coal Geology 70~(1-3) (2007) 184--192.

\bibitem{ayora2016recovery}
C.~Ayora, F.~Mac\`ias, E.~Torres, A.~Lozano, S.~Carrero, J.-M. Nieto,
  R.~P\'erez-L\`opez, A.~Fern\'andez-Martínez, H.~Castillo-Michel, Recovery of
  rare earth elements and yttrium from passive-remediation systems of acid mine
  drainage, Environmental Science \& Technology 50~(15) (2016) 8255--8262.

\bibitem{vital2018treatment}
B.~Vital, J.~Bartacek, J.~Ortega-Bravo, D.~Jeison, Treatment of acid mine
  drainage by forward osmosis: Heavy metal rejection and reverse flux of draw
  solution constituents, Chemical Engineering Journal 332 (2018) 85--91.

\bibitem{yan2013geochemistry}
Z.~Yan, G.~Liu, R.~Sun, Q.~Tang, D.~Wu, B.~Wu, C.~Zhou, Geochemistry of rare
  earth elements in groundwater from the taiyuan formation limestone aquifer in
  the wolonghu coal mine, anhui province, china, Journal of Geochemical
  Exploration 135 (2013) 54--62.

\bibitem{tian2020rare}
L.~Tian, H.~Chang, P.~Tang, T.~Li, X.~Zhang, S.~Liu, Q.~He, T.~Wang, J.~Yang,
  Y.~Bai, et~al., Rare earth elements occurrence and economical recovery
  strategy from shale gas wastewater in the sichuan basin, china, ACS
  Sustainable Chemistry \& Engineering 8~(32) (2020) 11914--11920.

\bibitem{voisin2017solubility}
T.~Voisin, A.~Erriguible, D.~Ballenghien, D.~Mateos, A.~Kunegel, F.~Cansell,
  C.~Aymonier, Solubility of inorganic salts in sub-and supercritical
  hydrothermal environment: Application to scwo processes, The Journal of
  Supercritical Fluids 120 (2017) 18--31.

\bibitem{yoon2021insitu}
T.~J. Yoon, J.~D. Riglin, P.~Sharan, R.~P. Currier, K.~A. Maerzke, A.~T.
  Findikoglu, An in-situ conductometric apparatus for physicochemical
  characterization of solutions and in-line monitoring of separation processes
  at elevated temperatures and pressures (2021).
\newblock \href {http://arxiv.org/abs/2109.14884} {\path{arXiv:2109.14884}}.

\bibitem{driesner2007system1}
T.~Driesner, C.~A. Heinrich, The system h2o--nacl. part i: Correlation formulae
  for phase relations in temperature--pressure--composition space from 0 to
  1000 c, 0 to 5000 bar, and 0 to 1 xnacl, Geochim. Cosmochim. Acta 71~(20)
  (2007) 4880--4901.

\bibitem{driesner2007system2}
T.~Driesner, The system h2o--nacl. part ii: Correlations for molar volume,
  enthalpy, and isobaric heat capacity from 0 to 1000 c, 1 to 5000 bar, and 0
  to 1 xnacl, Geochim. Cosmochim. Acta 71~(20) (2007) 4902--4919.

\bibitem{wagner2002iapws}
W.~Wagner, A.~Pru{\ss}, The iapws formulation 1995 for the thermodynamic
  properties of ordinary water substance for general and scientific use, J.
  Phys. Chem. Ref. Data 31~(2) (2002) 387--535.

\bibitem{zhang2010baseline}
Z.-M. Zhang, S.~Chen, Y.-Z. Liang, Baseline correction using adaptive
  iteratively reweighted penalized least squares, Analyst 135~(5) (2010)
  1138--1146.

\bibitem{savitzky1964smoothing}
A.~Savitzky, M.~J. Golay, Smoothing and differentiation of data by simplified
  least squares procedures., Analytical chemistry 36~(8) (1964) 1627--1639.

\bibitem{zhao2020graphite}
J.~Zhao, J.~H. Dumont, U.~Martinez, J.~Macossay, K.~Artyushkova, P.~Atanassov,
  G.~Gupta, Graphite intercalation compounds derived by green chemistry as
  oxygen reduction reaction catalysts, ACS Applied Materials \& Interfaces
  12~(38) (2020) 42678--42685.

\bibitem{gravzulis2012crystallography}
S.~Gra{\v{z}}ulis, A.~Da{\v{s}}kevi{\v{c}}, A.~Merkys, D.~Chateigner,
  L.~Lutterotti, M.~Quiros, N.~R. Serebryanaya, P.~Moeck, R.~T. Downs,
  A.~Le~Bail, Crystallography open database (cod): an open-access collection of
  crystal structures and platform for world-wide collaboration, Nucleic acids
  research 40~(D1) (2012) D420--D427.

\bibitem{allmann2007introduction}
R.~Allmann, R.~Hinek, The introduction of structure types into the inorganic
  crystal structure database icsd, Acta Crystallographica Section A:
  Foundations of Crystallography 63~(5) (2007) 412--417.

\bibitem{robinson2002electrolyte}
R.~A. Robinson, R.~H. Stokes, Electrolyte solutions, Courier Corporation, 2002.

\bibitem{zimmer2016supcrtbl}
K.~Zimmer, Y.~Zhang, P.~Lu, Y.~Chen, G.~Zhang, M.~Dalkilic, C.~Zhu, Supcrtbl: A
  revised and extended thermodynamic dataset and software package of supcrt92,
  Computers \& geosciences 90 (2016) 97--111.

\bibitem{sverjensky2014water}
D.~A. Sverjensky, B.~Harrison, D.~Azzolini, Water in the deep earth: the
  dielectric constant and the solubilities of quartz and corundum to 60 kb and
  1200 c, Geochimica et Cosmochimica Acta 129 (2014) 125--145.

\bibitem{shock1992calculation}
E.~L. Shock, E.~H. Oelkers, J.~W. Johnson, D.~A. Sverjensky, H.~C. Helgeson,
  Calculation of the thermodynamic properties of aqueous species at high
  pressures and temperatures. effective electrostatic radii, dissociation
  constants and standard partial molal properties to 1000 c and 5 kbar, Journal
  of the Chemical Society, Faraday Transactions 88~(6) (1992) 803--826.

\bibitem{shock1993metal}
E.~L. Shock, C.~M. Koretsky, Metal-organic complexes in geochemical processes:
  Calculation of standard partial molal thermodynamic properties of aqueous
  acetate complexes at high pressures and temperatures, Geochimica et
  Cosmochimica Acta 57~(20) (1993) 4899--4922.

\bibitem{zhang2005prediction}
Z.~Zhang, Z.~Duan, Prediction of the pvt properties of water over wide range of
  temperatures and pressures from molecular dynamics simulation, Physics of the
  Earth and Planetary Interiors 149~(3-4) (2005) 335--354.

\bibitem{fernandez1997formulation}
D.~Fernandez, A.~Goodwin, E.~W. Lemmon, J.~Levelt~Sengers, R.~Williams, A
  formulation for the static permittivity of water and steam at temperatures
  from 238 k to 873 k at pressures up to 1200 mpa, including derivatives and
  debye--h{\"u}ckel coefficients, Journal of Physical and Chemical Reference
  Data 26~(4) (1997) 1125--1166.

\bibitem{delany1978calculation}
J.~M. Delany, H.~C. Helgeson, Calculation of the thermodynamic consequences of
  dehydration in subducting oceanic crust to 100 kb and> 800 degrees c,
  American Journal of Science 278~(5) (1978) 638--686.

\bibitem{chan2021dewpython}
A.~Chan, M.~M. Daswani, S.~Vance, Dewpython: A python implementation of the
  deep earth water model and application to ocean worlds, arXiv preprint
  arXiv:2105.14096 (2021).

\bibitem{finney2019ion}
A.~R. Finney, S.~Lectez, C.~L. Freeman, J.~H. Harding, S.~Stackhouse, Ion
  association in lanthanide chloride solutions, Chemistry (Weinheim an der
  Bergstrasse, Germany) 25~(37) (2019) 8725.

\bibitem{stepanchikova1999spectrophotometric}
S.~Stepanchikova, G.~Kolonin, Spectrophotometric study of complexation of
  neodymium in chloride solutions at temperatures up to 250 c, Zh. Neorg. Khim
  44~(10) (1999) 1744--1751.

\bibitem{gammons1996aqueous}
C.~Gammons, S.~Wood, A.~Williams-Jones, The aqueous geochemistry of the rare
  earth elements and yttrium: Vi. stability of neodymium chloride complexes
  from 25 to 300 c, Geochimica et Cosmochimica Acta 60~(23) (1996) 4615--4630.

\bibitem{migdisov2002spectrophotometric}
A.~A. Migdisov, A.~Williams-Jones, A spectrophotometric study of neodymium
  (iii) complexation in chloride solutions, Geochimica et Cosmochimica Acta
  66~(24) (2002) 4311--4323.

\bibitem{haas1995rare}
J.~R. Haas, E.~L. Shock, D.~C. Sassani, Rare earth elements in hydrothermal
  systems: estimates of standard partial molal thermodynamic properties of
  aqueous complexes of the rare earth elements at high pressures and
  temperatures, Geochimica et Cosmochimica Acta 59~(21) (1995) 4329--4350.

\bibitem{wood1990aqueous}
S.~A. Wood, The aqueous geochemistry of the rare-earth elements and yttrium: 2.
  theoretical predictions of speciation in hydrothermal solutions to 350 c at
  saturation water vapor pressure, Chemical Geology 88~(1-2) (1990) 99--125.

\bibitem{migdisov2006spectrophotometric}
A.~A. Migdisov, V.~Reukov, A.~Williams-Jones, A spectrophotometric study of
  neodymium (iii) complexation in sulfate solutions at elevated temperatures,
  Geochimica et Cosmochimica Acta 70~(4) (2006) 983--992.

\bibitem{hnedkovsky2005electrical}
L.~Hnedkovsky, R.~H. Wood, V.~N. Balashov, Electrical conductances of aqueous
  na2so4, h2so4, and their mixtures: Limiting equivalent ion conductances,
  dissociation constants, and speciation to 673 k and 28 mpa, The Journal of
  Physical Chemistry B 109~(18) (2005) 9034--9046.

\bibitem{helgeson1974theoretical}
H.~C. Helgeson, D.~H. Kirkham, Theoretical prediction of the thermodynamic
  behavior of aqueous electrolytes at high pressures and temperatures; ii,
  debye-huckel parameters for activity coefficients and relative partial molal
  properties, American Journal of Science 274~(10) (1974) 1199--1261.

\bibitem{bromley1972approximate}
L.~A. Bromley, Approximate individual ion values of $\beta$ (or b) in extended
  debye-h{\"u}ckel theory for uni-univalent aqueous solutions at 298.15 k, The
  Journal of Chemical Thermodynamics 4~(5) (1972) 669--673.

\bibitem{pitzer1974thermodynamics}
K.~S. Pitzer, G.~Mayorga, Thermodynamics of electrolytes. iii. activity and
  osmotic coefficients for 2--2 electrolytes, Journal of Solution Chemistry
  3~(7) (1974) 539--546.

\bibitem{blum1975mean}
L.~Blum, Mean spherical model for asymmetric electrolytes: I. method of
  solution, Molecular Physics 30~(5) (1975) 1529--1535.

\bibitem{bernard1992conductance}
O.~Bernard, W.~Kunz, P.~Turq, L.~Blum, Conductance in electrolyte solutions
  using the mean spherical approximation, The Journal of Physical Chemistry
  96~(9) (1992) 3833--3840.

\bibitem{marcus1988ionic}
Y.~Marcus, Ionic radii in aqueous solutions, Chemical Reviews 88~(8) (1988)
  1475--1498.

\bibitem{sharygin2001tests}
A.~V. Sharygin, I.~Mokbel, C.~Xiao, R.~H. Wood, Tests of equations for the
  electrical conductance of electrolyte mixtures: measurements of association
  of nacl (aq) and na2so4 (aq) at high temperatures, The Journal of Physical
  Chemistry B 105~(1) (2001) 229--237.

\bibitem{powell1964efficient}
M.~J. Powell, An efficient method for finding the minimum of a function of
  several variables without calculating derivatives, The computer journal 7~(2)
  (1964) 155--162.

\bibitem{brassard2000feasible}
P.~Brassard, P.~Bodurtha, A feasible set for chemical speciation problems,
  Computers \& Geosciences 26~(3) (2000) 277--291.

\bibitem{richards1959flexible}
F.~Richards, A flexible growth function for empirical use, Journal of
  experimental Botany 10~(2) (1959) 290--301.

\bibitem{yoon2019electrical}
T.~J. Yoon, L.~A. Patel, M.~J. Vigil, K.~A. Maerzke, A.~T. Findikoglu, R.~P.
  Currier, Electrical conductivity, ion pairing, and ion self-diffusion in
  aqueous nacl solutions at elevated temperatures and pressures, The Journal of
  chemical physics 151~(22) (2019) 224504.

\bibitem{han2019effect}
K.~N. Han, Effect of anions on the solubility of rare earth element-bearing
  minerals in acids, Mining, Metallurgy \& Exploration 36~(1) (2019) 215--225.

\bibitem{maribo2013modeling}
B.~Maribo-Mogensen, G.~M. Kontogeorgis, K.~Thomsen, Modeling of dielectric
  properties of aqueous salt solutions with an equation of state, The Journal
  of Physical Chemistry B 117~(36) (2013) 10523--10533.

\bibitem{wang2001computation}
P.~Wang, A.~Anderko, Computation of dielectric constants of solvent mixtures
  and electrolyte solutions, Fluid Phase Equilibria 186~(1-2) (2001) 103--122.

\bibitem{patel2021nacl}
L.~A. Patel, T.~J. Yoon, R.~P. Currier, K.~A. Maerzke, Nacl aggregation in
  water at elevated temperatures and pressures: Comparison of classical force
  fields, The Journal of Chemical Physics 154~(6) (2021) 064503.

\bibitem{friesen2018hydration}
S.~Friesen, S.~Krickl, M.~Luger, A.~Nazet, G.~Hefter, R.~Buchner, Hydration and
  ion association of la 3+ and eu 3+ salts in aqueous solution, Physical
  Chemistry Chemical Physics 20~(13) (2018) 8812--8821.

\bibitem{van2019calcium}
A.~Van~Driessche, T.~Stawski, M.~Kellermeier, Calcium sulfate precipitation
  pathways in natural and engineered environments, Chemical Geology 530 (2019)
  119274.

\bibitem{palmer1986thermal}
D.~A. Palmer, S.~Drummond, Thermal decarboxylation of acetate. part i. the
  kinetics and mechanism of reaction in aqueous solution, Geochimica et
  Cosmochimica Acta 50~(5) (1986) 813--823.

\bibitem{bell1994thermal}
J.~L. Bell, D.~A. Palmer, H.~Barnes, S.~Drummond, Thermal decomposition of
  acetate: Iii. catalysis by mineral surfaces, Geochimica et cosmochimica acta
  58~(19) (1994) 4155--4177.

\bibitem{mccollom2003experimental}
T.~M. McCollom, J.~S. Seewald, Experimental study of the hydrothermal
  reactivity of organic acids and acid anions: Ii. acetic acid, acetate, and
  valeric acid, Geochimica et Cosmochimica Acta 67~(19) (2003) 3645--3664.

\bibitem{onwudili2010hydrothermal}
J.~A. Onwudili, P.~T. Williams, Hydrothermal reactions of sodium formate and
  sodium acetate as model intermediate products of the sodium
  hydroxide-promoted hydrothermal gasification of biomass, Green chemistry
  12~(12) (2010) 2214--2224.

\bibitem{ong2013situ}
A.~Ong, J.~Pironon, P.~Robert, J.~Dubessy, M.-C. Caumon, A.~Randi, O.~Chailan,
  J.-P. Girard, In situ decarboxylation of acetic and formic acids in aqueous
  inclusions as a possible way to produce excess ch 4, Geofluids 13~(3) (2013)
  298--304.

\bibitem{li2017experimental}
Y.~Li, S.~Zhou, J.~Li, Y.~Ma, K.~Chen, Y.~Wu, Y.~Zhang, Experimental study of
  the decomposition of acetic acid under conditions relevant to deep
  reservoirs, Applied Geochemistry 84 (2017) 306--313.

\bibitem{mochizuki1974synthesis}
A.~Mochizuki, K.~Nagashima, H.~Wakita, The synthesis of crystalline hydrated
  double carbonates of rare earth elements and sodium, Bulletin of the Chemical
  Society of Japan 47~(3) (1974) 755--756.

\bibitem{rao1999solubility}
L.~Rao, D.~Rai, A.~R. Felmy, C.~F. Novak, Solubility of nand (co 3) 2. 6h 2 o
  (c) in mixed electrolyte (na-cl-co 3-hco 3) and synthetic brine solutions,
  in: Actinide Speciation in High Ionic Strength Media, Springer, 1999, pp.
  153--169.

\bibitem{yang2019metastable}
Y.~Yang, X.~Zhang, L.~Li, T.~Wei, K.~Li, Metastable dissolution regularity of
  nd3+ in na2co3 solution and mechanism, ACS omega 4~(5) (2019) 9160--9168.

\bibitem{shaplygin1979thermogravimetric}
I.~Shaplygin, V.~Komarov, V.~Lazarev, A thermogravimetric study of praseodymium
  (iii), neodymium, samarium, gadolinium and holmium acetates, benzoates and
  abietates, Journal of thermal analysis 15~(2) (1979) 215--223.

\bibitem{kkepinski2004hydrothermal}
L.~K{\c{e}}pi{\'n}ski, M.~Zawadzki, W.~Mi{\'s}ta, Hydrothermal synthesis of
  precursors of neodymium oxide nanoparticles, Solid state sciences 6~(12)
  (2004) 1327--1336.

\bibitem{bukin1972neutron}
V.~Bukin, Neutron diffraction-and x-ray diffraction study of the crystalline
  structure of neodymium hydro xyl-chloride, nd(oh)\_2cl, in: Doklady Akademii
  Nauk, Vol. 207, Russian Academy of Sciences, 1972, pp. 1332--1335.

\bibitem{zehnder2010investigation}
R.~A. Zehnder, D.~L. Clark, B.~L. Scott, R.~J. Donohoe, P.~D. Palmer, W.~H.
  Runde, D.~E. Hobart, Investigation of the structural properties of an
  extended series of lanthanide bis-hydroxychlorides ln (oh) 2cl (ln= nd- lu,
  except pm and sm), Inorganic chemistry 49~(11) (2010) 4781--4790.

\bibitem{gomez2008anhydrous}
S.~Gomez~Torres, G.~Meyer, Anhydrous neodymium (iii) acetate, Zeitschrift
  f{\"u}r anorganische und allgemeine Chemie 634~(2) (2008) 231--233.

\bibitem{beall1976refinement}
G.~Beall, W.~Milligan, D.~Dillin, R.~Williams, J.~McCoy, Refinement of
  neodymium trihydroxide, Acta Crystallographica Section B: Structural
  Crystallography and Crystal Chemistry 32~(7) (1976) 2227--2229.

\bibitem{dexpert1974determination}
H.~Dexpert, P.~Caro, Determination de la structure cristalline de la variete a
  des hydroxycarbonates de terres rares lnohco3 (ln=nd), Materials Research
  Bulletin 9~(11) (1974) 1577--1585.

\bibitem{christensen1970hydrothermal}
A.~N. Christensen, A.~N{\o}rlund, Hydrothermal preparation of neodymium oxide
  carbonate. the location of the carbonate ion in the structure of nd2o2co3,
  Acta Chem Scand 24 (1970) 2440--2446.

\end{thebibliography}
\end{document}